\documentclass[intlimits,twoside,a4paper]{article}
\usepackage{amstext}
\usepackage{graphicx}
\usepackage{color}
\usepackage{amssymb}
\usepackage{amsmath}
\usepackage[T2A]{fontenc}
\usepackage[cp1251]{inputenc}
\usepackage[eqsecnum]{cmpj}
\issue{2011}{14}{3}{33605}

\doinumber{10.5488/CMP.14.33605}

\title[Maier-Saupe nematogenic fluid: field theoretical approach]{Maier-Saupe nematogenic fluid: field theoretical approach}

\author[M. Holovko, D. di Caprio, I. Kravtsiv]
{ M. Holovko\refaddr{label1}, D. di Caprio\refaddr{label2}, I.
Kravtsiv\refaddr{label1}\thanks{E-mail: ivankr@icmp.lviv.ua}}

\addresses{
\addr{label1} Institute for Condensed Matter Physics of the
National Academy of Sciences of Ukraine, \\1~Svientsitskii Str.,
79011~Lviv, Ukraine
\addr{label2} Laboratoire d'Electrochimie, Chimie des Interfaces
et Mod\'elisation pour l'Energie (LECIME) ENSCP, \\ Chimie
ParisTech, Case~39, 4~Pl.~Jussieu, 75005~Paris, France}

\date{Received June 29, 2011, in final form August 4, 2011}

\authorcopyright{M. Holovko, D. di Caprio, I. Kravtsiv, 2011}

\begin{document}
\maketitle

\begin{abstract}
We adopt a field theoretical approach to the study of the structure
and thermodynamics of a homogeneous Maier-Saupe nematogenic
fluid interacting with anisotropic Yukawa potential. In the
mean field approximation we retrieve a standard Maier-Saupe
theory for liquid crystals. In this theory, the single-particle distribution function is
expressed via the second order Legendre polynomial of molecule
orientations. In the Gaussian approximation we obtain
analytical expressions for correlation functions, free
energy, pressure, chemical potential, and
elasticity constant. Subsequently we find corrections due to
fluctuations and show that the single-particle distribution function now contains Legendre
polynomials of higher orders. We also use Ward symmetry
identities to set a simple condition for correlation
functions.



\end{abstract}

\section{Introduction}
Maier-Saupe nematogenic fluid~\cite{masa} is one of the
simplest models that account for the isotropic-nematic phase
transition in the liquid crystal phase. The properties of this
model have been intensively studied by the liquid theory
methods such as integral equations for correlation functions~\cite{hoso,hoso1,pe,misi,loma,soho}. In the integral equation
theory there is a problem of the correctness of taking the
fluctuation effects into account, the treatment of which
depends on closure relations used in integral equations. In
order to treat the fluctuations more properly and to control
the level of this treatment, in this paper we will apply the
field theoretical approach. This is the first time the field
theoretical approach is applied to the description of
anisotropic molecular fluids.

The method we are proposing focuses on fluctuations of the
field at a given point and implements a perturbative scheme by
expanding the Hamiltonian on density fluctuations. In the past,
the statistical field theory proved to be successful in the
description of a variety of systems with Coulomb~\cite{castaba,ionic1,ionic2,ionic3,ionic4} and Yukawa-type
interactions~\cite{molphys,castaba2}. In this work we show that
this approach also reproduces the familiar results for
anisotropic systems, notably the mean field Maier-Saupe theory.
Subsequently we go beyond this approximation and obtain an
analytical expression for the pair correlation function.
% Our results predict the appearance of Goldstone modes in the system
% which is in full agreement with the theory of de Gennes
%~\cite{ge}.
In the Gaussian approximation we also obtain new results for
the main structural and thermodynamic properties of the system.
The expressions we derive contain the orientational order
parameter allowing us to compare the results for the isotropic
and nematic phases. Finally, we calculate the correction to the
mean field single-particle distribution function due to fluctuations which is expressed in
terms of the fourth order Legendre polynomials of molecule
orientations. Our results for the pair correlation functions
predict the appearance of Goldstone modes in the system which
is in full agreement with the theory of de Gennes~\cite{ge}.

\newpage

For the purpose of simplification, in this paper we consider a
fluid of point particles. However, in the future we hope to
modify the obtained results for non-point particles using the
mean spherical results~\cite{hoso,hoso1,soho} as it was done
for a non-point ionic system~\cite{hoio}.

\section{The model and field theory formalism}
We consider a molecular fluid of particles interacting via an
anisotropic Yukawa-type potential $\nu(r_{12}\,,\Omega_1\Omega_2)$:
\begin{align}
\label{potential}
\nu(r_{12}\,,\Omega_1\Omega_2)&=\frac{A}{r_{12}}\re^{-\alpha
r_{12}}P_2(\cos{\theta_{12}})\nonumber\\
&=\frac{A}{r_{12}}\re^{-\alpha
r_{12}}\frac{1}{5}\sum\limits_{m}{Y^*_{2m}(\Omega_1)Y_{2m}(\Omega_2)},
\end{align}
where $r_{12}$ denotes the distance between particles 1 and 2,
$\Omega=\left(\theta,\phi\right )$ are orientations of
particles, $P_2(\cos{\theta_{12}})=(3\cos^2\theta_{12}-1)/2$ is
the second order Legendre polynomial of relative molecule
orientations, $Y_{lm}(\Omega)$ are standard spherical harmonics
\cite{gra} without the normalization factor $1/\sqrt{4\pi}$,
$A$ is the amplitude of the interaction, and $\alpha$ is the
inverse range.

In a series of papers on ionic and Yukawa fluids~\cite{castaba,castaba2,molphys} it was shown that it is
possible to describe these fluids using the field theoretical
approach. In this paper we will develop this approach for the
description of an anisotropic molecular fluid with the
interaction of the form~(\ref{potential}).

Within the field-theoretical formalism, the Hamiltonian is a
functional of density field and can be written as
    \begin{align}
    \beta H[\rho(\mathbf{r},\Omega)]&=\beta
    H^{\rm entr}[\rho(\mathbf{r},\Omega)]+\beta
    H^{\rm int}[\rho(\mathbf{r},\Omega)]\nonumber\\
    &=\int\rho(\mathbf{r},\Omega)\left[\ln(\rho(\mathbf{r},\Omega)\Lambda_{\rm R}\Lambda_{\rm T}^3)-1\right]\rd{\mathbf{r}\rd\Omega}\nonumber\\
    &+\frac{\beta}{2}\int{\nu(r_{12}\,,\Omega_1\Omega_2)\rho(\mathbf{r}_1\,,\Omega_1)
    \rho(\mathbf{r}_2\,,\Omega_2)}
    \rd{\mathbf{r}_1}\rd{\mathbf{r}_2\rd{\Omega_1}\rd{\Omega_2}}\,,
    \end{align}
where $\beta=1/k_{\rm B}T$ is the inverse temperature,
$\rd\Omega=(1/4\pi)\sin\theta \rd\theta \rd\phi$ is the normalized
angle element, $\rho(\mathbf{r},\Omega)$ is particle density
per angle such that
$\int{\rho(\mathbf{r},\Omega)\rd{\Omega}}=\rho(\mathbf{r})$,
$\Lambda_{\rm T}$ is the thermal de Broglie wavelength of the
molecules, and the quantity $\Lambda^{-1}_{\rm R}$ is the rotational
partition function for a single molecule~\cite{gra}.

As in previous papers~\cite{castaba,castaba2,molphys}, we adopt
the canonical ensemble approach. We fix the number of particles
by the condition $\int \rho(\mathbf{r})\rd\mathbf{r}=N$ or
$\frac{1}{V}\int \rho(\mathbf{r})\rd\mathbf{r}=\rho$, where $V$
is the volume and $\rho$ is the average density of the system.
To verify this condition in a formally unconstrained calculus
we introduce a Lagrange multiplier $\lambda$ such that
\begin{equation}
\label{MFA1}
\frac{\delta\beta
H[\rho(\mathbf{r},\Omega)]}{\delta\rho(\mathbf{r},\Omega)}=\lambda.
\end{equation}
The partition function
$Z_N\left[\rho(\mathbf{r},\Omega)\right]$ can be expressed as
\begin{align}
Z_N\left[\rho(\mathbf{r},\Omega)\right]=\int\textit{D}\rho(\mathbf{r},\Omega)\exp\{-\beta
H[\rho(\mathbf{r},\Omega)]\},\nonumber
\end{align}
where $\textit{D}\rho(\mathbf{r},\Omega)$ denotes functional
integration over all possible density distributions such that
the total number of particles is $N$. The logarithm of the
partition function gives the Helmholtz free energy
\begin{align}
\beta F=-\ln Z_N\,.
\end{align}
Due to the character of the interparticle interaction, the
considered system is characterized by two non-dimensional
parameters: non-dimensional density $\rho^*=\rho/\alpha^3$ and
non-dimensional inverse temperature $\beta^*=-\beta
A\alpha=1/T^*$. As we will see in our calculations, the third
non-dimensional parameter appears $M^*=-4\pi\rho\beta
A/\alpha^2=4\pi\rho^*\beta^*$.

In order to calculate the functional integral, we expand the
Hamiltonian around the real angle-dependent density $\rho(\Omega)$ which in the
homogeneous case does not depend on $\mathbf{r}$:
\begin{align}
\label{ham_exp}
\beta
H&[\rho(\Omega)+\delta\rho(\mathbf{r},\Omega)]=\nonumber\\
\nonumber &=\int\big(\rho(\Omega)+\delta\rho(\mathbf{r},\Omega)\big)
\left[\ln\left(\frac{\rho(\Omega)}{\rho}\right)+
\ln\left(1+\frac{\delta\rho(\mathbf{r},\Omega)}{\rho(\Omega)}\right)-1\right]
\rd\mathbf{r}\rd\Omega\nonumber\\
    &+\frac{\beta}{2}\int{\nu(r_{12}\,,\Omega_1\Omega_2)\big(\rho(\Omega_1)+\delta\rho(\mathbf{r}_1\,,\Omega_1)\big)
    \big(\rho(\Omega_2)+\delta\rho(\mathbf{r}_2\,,\Omega_2)\big)}
    \rd{\mathbf{r}_1}\rd{\mathbf{r}_2\rd{\Omega_1}\rd{\Omega_2}}\,.
\end{align}

\section{Mean field approximation}
In order to obtain thermodynamic properties of the considered
fluid we need to calculate the partition function. The lowest
order approximation for the partition function is the saddle
point for the functional integral which is the mean field
approximation (MFA) from the physical point of view. In the
canonical formalism it corresponds to fixing the Lagrange
parameter $\lambda$ such that the relation~(\ref{MFA1}) is true
for the average density.

Expanding the logarithm in~(\ref{ham_exp}), we obtain
\begin{equation}
\label{A}
 \frac{\delta\beta
H[\rho(\mathbf{r},\Omega)]}{\delta\rho(\mathbf{r}_1\,,\Omega)}=\ln\frac{\rho(\Omega)}{\rho}+
\beta\int\nu(r_{12}\,,\Omega_1\Omega_2)\rho(\Omega_2)\rd{\mathbf{r}_2}\rd{\Omega_2}\,.
\end{equation}
The second term  on the right-hand side of equation~(\ref{A}) equals
\begin{align}
\label{expanding}
&\beta\int\nu(r_{12}\,,\Omega_1\Omega_2)\rho(\Omega_2)\rd{\mathbf{r}_2}\rd{\Omega_2}=\nonumber\\
&=\beta\int\frac{A}{r_{12}}\re^{-\alpha
r_{12}}\rd{\mathbf{r}_2}\int\frac{1}{5}\sum\limits_{m'}{Y^*_{2m'}(\Omega_2)Y_{2m'}(\Omega_1)}
\sum\limits_{lm''}\rho_{lm''}Y_{lm''}(\Omega_2)\rd{\Omega_2}\nonumber\\
&=\beta\frac{1}{5}\nu\sum\limits_{m}\rho_{2m}Y_{2m}(\Omega_1),
\end{align}
where we have used
\begin{align}
\nu &=\int\frac{A}{r_{12}}\re^{-\alpha
r_{12}}\rd{\mathbf{r}_{12}}=\frac{4\pi A}{\alpha^2}\,.
\end{align}
If we choose the value of parameter $\lambda$ to be
\begin{equation}
\re^{\lambda}\equiv \left(\int{\rd}\Omega \exp\left[-
\beta\frac{1}{5}\nu\sum\limits_{m}\rho_{2m}Y_{2m}(\Omega)\right]\right)^{-1}
\equiv \frac{1}{Z}
\end{equation}
then from (\ref{A}) we get the following equation for density
within MFA:
\begin{align}
\label{B}
 \rho(\Omega)=\frac{\rho}{Z}\,\exp\left[-
\beta\frac{1}{5}\nu\sum\limits_{m}\rho_{2m}Y_{2m}(\Omega)\right]\equiv \rho f(\Omega),
\end{align}
where $f(\Omega)$ is the single-particle distribution function
and the averages can be calculated according to $\langle \ldots
\rangle_{\Omega}=\int f(\Omega)(\ldots)\rd\Omega$.

%\begin{figure}
%\centerline{\includegraphics[scale=0.45]{graph5}}
%\caption {Dependence of orientational order parameter S on parameter $1/\mathrm{M}^*$}
%\label{buka1}
%\end{figure}
%\begin{figure}
%\centerline{\includegraphics[scale=0.45]{phase_diagram}}
%\caption{Density-temperature phase diagram }
%%\newline on parameter 1/M^*} At $\mathbf{\rho_b\nu_2=0.433}$ S=0.675.}
%\label{buka2}
%\end{figure}

\begin{figure}[!h]
\vspace{-0.5cm}
\hspace{-1cm}
\includegraphics[width=0.6\textwidth]{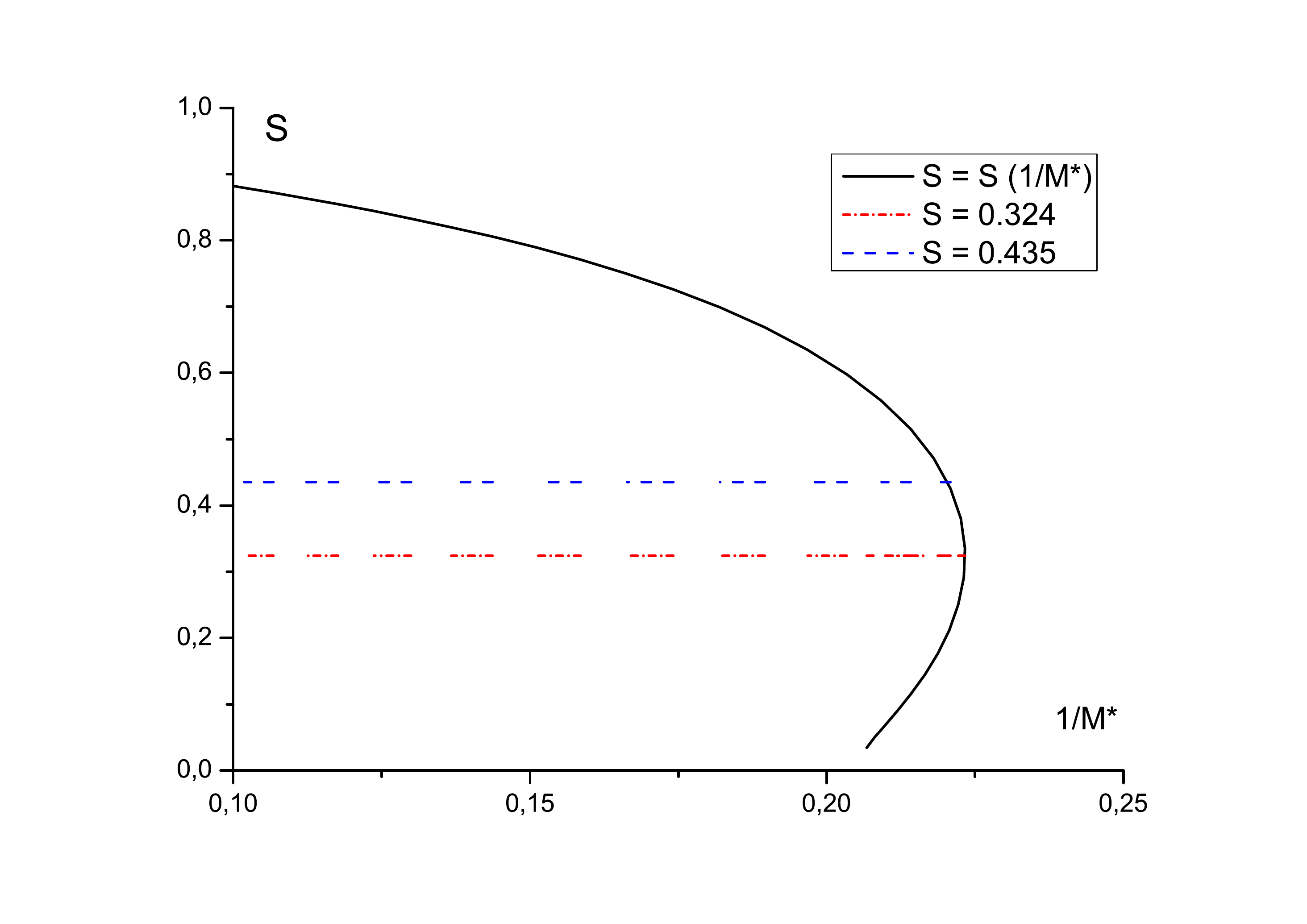}%
\hfill%
\hspace{-1.5cm}
\includegraphics[width=0.6\textwidth]{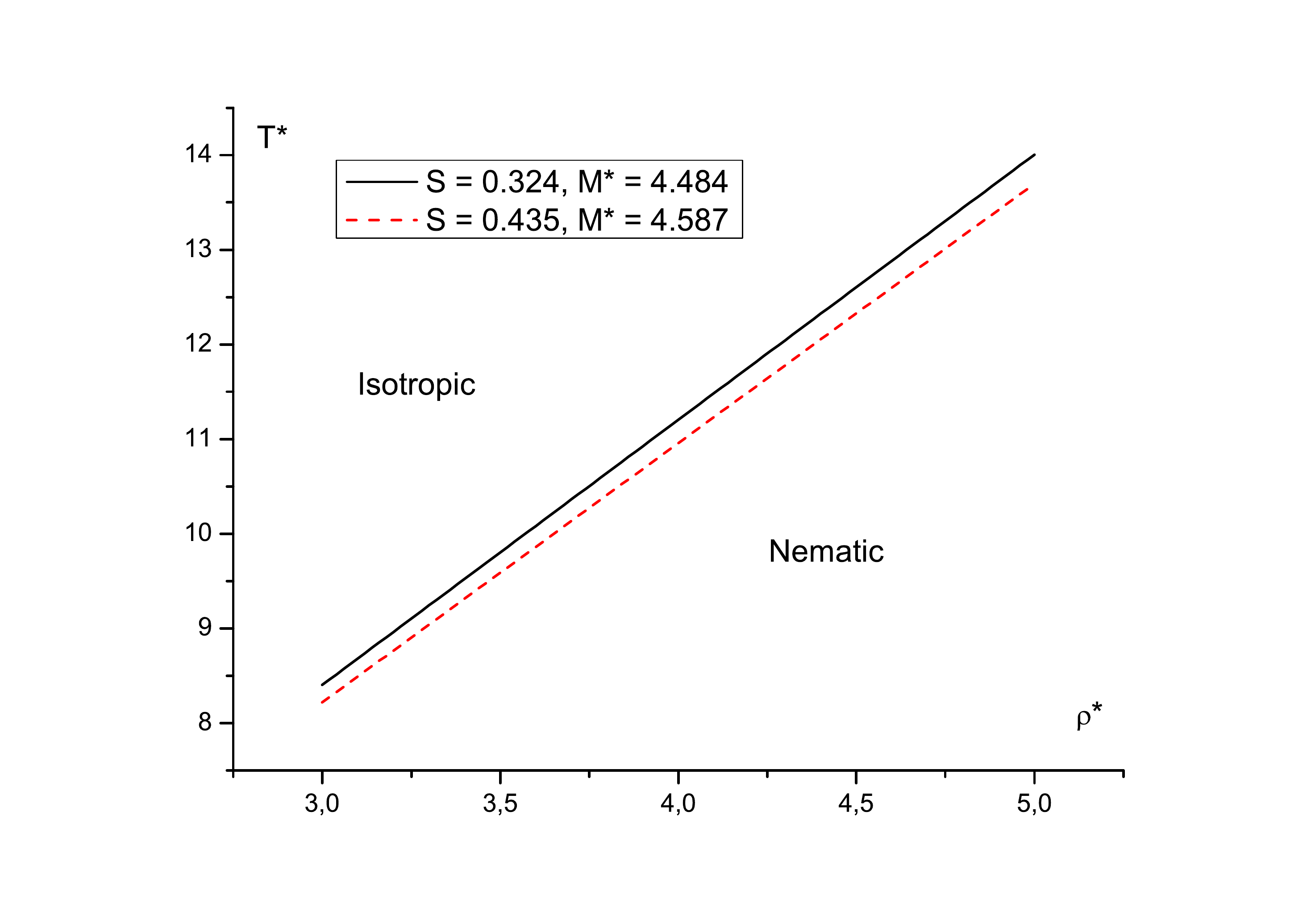}%
\vspace{-1cm}
\\
\parbox[t]{0.5\textwidth}{%
\caption {Dependence of orientational order parameter S on parameter $1/{M}^*$.}\label{buka1}}%
\hfill%
\parbox[t]{0.5\textwidth}{%
\caption{Density-temperature phase diagram.}
\label{buka2}
}%
\end{figure}

If we multiply both sides of equation~(\ref{B}) by $Y_{2m}(\Omega)$
and integrate by $\rd\Omega$  we will obtain
\begin{align}
\label{2m}
\sum\limits_{l'm'}\rho_{l'm'} \int \rd{\Omega}\,Y_{l'm'}(\Omega)Y_{2m}(\Omega)&=
\rho\langle Y_{2m}(\Omega)\rangle_{\Omega}\,,\nonumber\\
\rho_{2m}&=\rho\langle Y_{2m}(\Omega)\rangle_{\Omega}\,.
\end{align}
In normal nematics, the orientational distribution function
$f(\Omega)$ is axially symmetric with respect to a preferred
direction $\mathbf{n}$ and depends only on the angle $\theta$
between the molecular orientation $\Omega$ and $\mathbf{n}$~\cite{hoso}. This means that only quantities independent of
angle $\phi$, in the plane perpendicular to $\mathbf{n}$, yield
non-zero averages and therefore for any $m\ne 0$ the averages
$\langle Y_{2m}(\Omega)\rangle_{\Omega}$ equal 0. As a result,
we obtain a well-known Maier-Saupe equation~\cite{ge}
\begin{align}
\label{maier}
\langle
Y_{20}(\Omega)\rangle_{\Omega}&=\frac{1}{Z}\,\int
Y_{20}(\Omega)\exp\left[ - \beta\frac{1}{5}\nu\rho\langle
Y_{20}(\Omega)\rangle_{\Omega} Y_{20}(\Omega)\right]\rd\Omega.
\end{align}
In terms of the orientational order parameter $S$ and reduced
unit $M^*$, equations~(\ref{B}) and~(\ref{maier}) can be
rewritten as follows:
\begin{align}
\label{density_reduced}
&\frac{\rho(\Omega)}{\rho}=\frac{\exp\left[\frac{3}{2}M^*S\cos^2\theta\right]}
{\int\limits_0^1  \exp\left[\frac{3}{2}M^*Sx^2\right]\rd x}\,,\\
\label{order}
&S=\langle{P_2(\cos{\theta})}\rangle_{\Omega}=\frac{1}{\sqrt{5}}\langle Y_{20}(\Omega)\rangle_{\Omega}=
-\frac{1}{2}+\frac{3}{2}\,
\frac{ \int\limits_0^1 x^2\exp\left[\frac{3}{2}M^*S x^2\right]\rd x}
{ \int\limits_0^1  \exp\left[\frac{3}{2}M^*Sx^2\right]\rd x}\,.
\end{align}

The order parameter $S$ can take on values from 0 to 1 with
values $S>0$ corresponding to the nematic phase. Equation~(\ref{order})
is self-consistent and must be solved
numerically. The resulting relationship between $S$ and $M^*$
is presented in figure~\ref{buka1}. The theory predicts a weak
first-order phase transition from the isotropic phase with
$S=0$ to the nematic phase with $S>0$. The smallest value of
the order parameter $S=0.324$ corresponding to $M^*=4.484$
defines the stability of the isotropic phase. A stable nematic
phase is given by the solution that minimizes the free energy
which in the MFA can be presented in the form
\begin{align}
\frac{\beta F}{N}=\ln\left(\rho\Lambda_{\rm T}^3\Lambda_{\rm R}\right)-1+\frac{1}{2}M^*S^2.
\end{align}
As a result, the stable nematic phase appears at $M^*=4.587$
and the value of the order parameter at the transition is
$S=0.435$. The region between $M^*=4.484$ and $M^*=4.587$
corresponds to the two-phase region which separates the
isotropic and the nematic phases. The corresponding phase
diagram in ``density-temperature'' coordinates is presented in
figure~\ref{buka2}.

\section{Fluctuation and correlation effects: Gaussian approximation}
In the MFA, fluctuations are neglected. In this section we take
them into account. To this end we should expand the
Hamiltonian. From~(\ref{ham_exp}), the quadratic term in the
Hamiltonian equals
\begin{align}
\beta H_2\left[\rho(\mathbf{r},\Omega)\right]&=\frac{1}{2}\int\frac{1}{\rho(\Omega)}\,\delta\rho^2(\mathbf{r},\Omega)
\rd{\mathbf{r}}\rd\Omega\nonumber\\
&+\frac{\beta}{2}\int\nu(r_{12}\,,\Omega_1\Omega_2)\delta\rho(\mathbf{r}_1\,,\Omega_1)\delta\rho(\mathbf{r}_2\,,\Omega_2)
\rd{\mathbf{r}_1}\rd{\mathbf{r}_2}\rd\Omega_1 \rd\Omega_2\,.
\end{align}
Expanding on the Fourier components
\begin{align}
\label{fourier}
\delta\rho(\mathbf{r},\Omega)
=\sum\limits_\mathbf{k}\delta\rho(\mathbf{k},\Omega)\re^{\ri\mathbf{kr}},
\end{align}
we obtain the expression for the quadratic term in the
$\mathbf{k}$-space
\begin{align}
\beta H_2\left[\rho(\mathbf{k},\Omega)\right]&=\frac{V}{2}\sum\limits_{\mathbf{k}}\int
\rd\Omega_1 \rd\Omega_2\delta\rho(\mathbf{k},\Omega_1)\delta\rho(\mathbf{-k},\Omega_2)\nonumber\\
&\times\left(\frac{\delta_{\Omega_1\Omega_2}}{\rho(\Omega_1)}+\frac{4\pi\beta
A}{k^2+\alpha^2}\frac{1}{5}\sum\limits_{m}{Y^*_{2m}(\Omega_1)Y_{2m}(\Omega_2)}\right).
\end{align}
\subsection{Correlation functions}
The expression for a pair correlation function $h$
is
\begin{align}
\label{corr1}
h(r_{12}\,,\Omega_1\Omega_2)\langle\rho(\mathbf{r}_1\,,\Omega_1)\rangle\,
\langle\rho(\mathbf{r}_2\,,\Omega_2)\rangle &=
\langle\delta\rho(\mathbf{r}_1\,,\Omega_1)\delta\rho(\mathbf{r}_2\,,\Omega_2)\rangle
\nonumber\\
- \langle\delta\rho(\mathbf{r}_1\,,\Omega_1)\rangle\,
\langle\delta\rho(\mathbf{r}_2\,,\Omega_2)\rangle
&-\delta(\mathbf{r}_1 -\mathbf{r}_2)\delta_{\Omega_1\Omega_2}\langle\rho(\mathbf{r}_1\,,\Omega_1)\rangle.
\end{align}
The second term on the right-hand side of equation~(\ref{corr1})
disappears like in the homogeneous case
$\langle\delta\rho(\mathbf{r},\Omega)\rangle=0$. The first term
equals
\begin{align}
\langle\delta\rho(\mathbf{r}_1\,,\Omega_1)\delta\rho(\mathbf{r}_2\,,\Omega_2)\rangle &=\frac{1}{Z_N}\int\textit{D}
\left(\delta\rho(\mathbf{r},\Omega)\right)\,\re^{
-\beta H_2\left[\rho(\mathbf{r},\Omega)\right]}
\delta\rho(\mathbf{r}_1\,,\Omega_1)\delta\rho(\mathbf{r}_2\,,\Omega_2)\nonumber\\
&=\sum\limits_{\mathbf{k}}\re^{\ri\mathbf{k}(\mathbf{r}_1-\mathbf{r}_2)}\langle\delta\rho(\mathbf{k},\Omega_1)
\delta\rho(\mathbf{-k},\Omega_2)\rangle\,
\nonumber\\
\label{gaussian}
&=\sum\limits_{\mathbf{k}}\re^{\ri\mathbf{k}(\mathbf{r}_1-\mathbf{r}_2)}\frac{ \int\textit{D}
\left(\delta\rho(\mathbf{k},\Omega)\right)\,\re^{
-\beta H_2\left[\rho(\mathbf{k},\Omega)\right]}
\delta\rho(\mathbf{k},\Omega_1)\delta\rho(\mathbf{-k},\Omega_2)}{ \int\textit{D}
\left(\delta\rho(\mathbf{k},\Omega)\right)\,\re^{
-\beta H_2\left[\rho(\mathbf{k},\Omega)\right]}}\,.
\end{align}
As in the basis~(\ref{fourier}) the Hamiltonian is of diagonal
form, and the Gaussian integral~(\ref{gaussian}) yields
\begin{align}
\label{O1}
\langle\delta\rho(\mathbf{k},\Omega_1)\delta\rho(\mathbf{-k},\Omega_2)\rangle=
\frac{1}{V}\left(\frac{\delta_{\Omega_1\Omega_2}}{\rho(\Omega_1)}+\frac{4\pi\beta
A}{k^2+\alpha^2}\frac{1}{5}\sum\limits_{m}{Y^*_{2m}(\Omega_1)Y_{2m}(\Omega_2)}\right)^{-1}.
\end{align}
The inverse of the matrix in brackets is
\begin{align}
\label{O2}
\left(\frac{\delta_{\Omega_1\Omega_2}}{\rho(\Omega_1)}+\frac{4\pi\beta
A}{k^2+\alpha^2}\frac{1}{5}\sum\limits_{m}{Y^*_{2m}(\Omega_1)Y_{2m}(\Omega_2)}\right)^{-1}
=%\nonumber\\
h(k,\Omega_1\Omega_2)\rho(\Omega_1)\rho(\Omega_2)+\delta_{\Omega_1\Omega_2}\rho(\Omega_1).
\end{align}
Identity (\ref{O2}) is in essence the Ornstein-Zernike
equation in the random phase approximation (RPA) for point
particles~\cite{yuho,hansen}
\begin{align} \label{Omain}
h(k,\Omega_1\Omega_2)=C(k,\Omega_1\Omega_2)+\int \rd\Omega_3
C(k,\Omega_1\Omega_3)h(k,\Omega_2\Omega_3)\rho(\Omega_3)
\end{align}
with the closure
\begin{align}
\label{closure}
C(\mathbf{r}_{12}\,,\Omega_1\Omega_2)=-\beta\nu(r_{12}\,,\Omega_1\Omega_2),
\end{align}
where $C(k,\Omega_1\Omega_2)$ and $h(k,\Omega_1\Omega_2)$ are
the Fourier transforms of the direct and pair correlation
functions, respectively.

In (\ref{Omain}) we can expand $f(k,\Omega_1\Omega_2)$ on
spherical harmonics
\begin{align}
f(k,\Omega_1\Omega_2)=\sum\limits_{lmnn'}f_{lmnn'}(k)Y^*_{lm}(\Omega_1)Y_{nn'}(\Omega_2).
\end{align}
Due to the closure (\ref{closure}) and symmetry properties of
the nematic we can write $f(r_{12}\,,\Omega_1\Omega_2)$ in the
form
\begin{align}
f(r_{12}\,,\Omega_1\Omega_2)&=\sum\limits_{m}f_{22m}(r_{12})Y_{2m}^*(\Omega_1)Y_{2m}(\Omega_2),
\end{align}
where
\begin{align}
C_{22m}(r)&=-\frac{1}{5}\frac{\beta A}{r}\re^{  -\alpha r}.
\end{align}
This reduces to the following equation for harmonics
\begin{align}
h_{22m}(k)=C_{22m}(k)+\langle Y_{2m}^2(\Omega)\rangle_{\Omega}\,\rho h_{22m}(k)C_{22m}(k),
\end{align}
resulting in the following expression for the harmonics of a
pair correlation function
\begin{align}
h_{22m}(k)=-\frac{1}{5}\frac{4\pi\beta A}{k^2+\alpha^2+
\langle Y_{2m}^2(\Omega)\rangle_{\Omega}\,\frac{1}{5}4\pi\rho\beta  A}\,,
\end{align}
which is a renormalized, ``effective'' Yukawa potential in the
{\bf{k}}-space. In the {\bf{r}}-space
\begin{align}
h_{22m}(r)=-\frac{1}{5}\frac{\beta A}{r}\,\exp\left[  -r\sqrt{\alpha^2+
\langle Y_{2m}^2(\Omega)\rangle_{\Omega} \frac{1}{5}4\pi\rho\beta  A}\right],
\end{align}
where $\langle Y^2_{2m}(\Omega)\rangle_{\Omega}=(1/\rho)\int
\rd\Omega \rho(\Omega)\vert Y_{2m}^2(\Omega)\vert$.

\begin{figure}
\centerline{\includegraphics[scale=0.35]{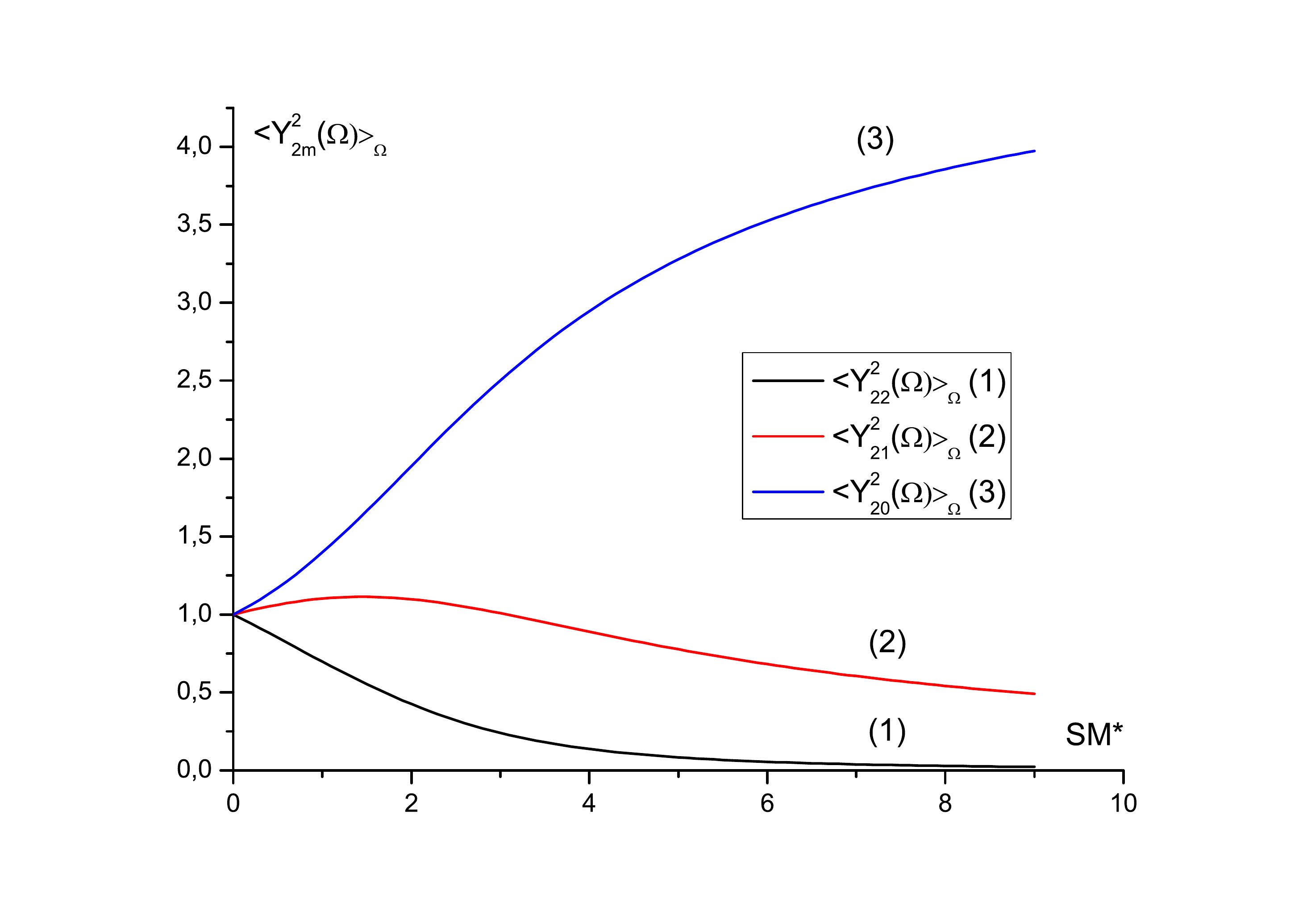}}
\caption{Dependence of quantities $\langle Y_{2m}^2(\Omega)\rangle_{\Omega}$
 on parameter $SM^*$ for $m=0,1,2$.}
\label{buka3}
\end{figure}

Dependencies of quantities $\langle
Y^2_{2m}(\Omega)\rangle_{\Omega}$ on the product $S M^*$ are
presented in figure~\ref{buka3}. Due to the normalization
condition of functions $Y_{2m}(\Omega)$, for $SM^*$=0, the
averages $\langle Y^2_{2m}(\Omega)\rangle_{\Omega}=1$.

\subsection{Correction to the single-particle distribution function}
Correction to the single-particle distribution function due to
Gaussian fluctuations can be found according to
\begin{align}
\frac{\rho^{\rm CR}(\Omega)}{\rho}&=\frac{f^{\rm MF}(\Omega)}{Z'}\big(1+[h(r,\Omega_1\Omega_2)-C(r,\Omega_1\Omega_2)]\big)
\left|_{
\begin{smallmatrix}
\Omega_1\rightarrow\Omega_2\\\mathbf{r}_1\rightarrow\mathbf{r}_2
\end{smallmatrix}}\right.\nonumber\\
%\mathop{\bigg\vert_{\Omega_1\rightarrow\Omega_2}}
%\limits_{\mathbf{r}_1\rightarrow\mathbf{r}_2}\\
&=\frac{1}{Z'}\,\re^{  \,
\frac{3}{2}M^*S\cos^2\theta}\big(1+\sum\limits_m
Y_{2m}^*(\Omega)Y_{2m}(\Omega)[h_{22m}(r)+\beta\nu(r)]\big)\bigg\vert_{r\rightarrow
0},
\end{align}
where the normalization constant $Z'$ can be found from
condition $\int\rho^{\rm CR}(\Omega)\rd\Omega=\rho$. Since
\begin{align}
\lim_{r\rightarrow 0}[h_{22m}(r)+\beta\nu(r)]=-
\frac{1}{5}\beta^* \left[\sqrt{1-\frac{1}{5}\langle Y_{2m}^2(\Omega)\rangle_{\Omega}
M^*}-1\right],
\end{align}
then the corrected single-particle distribution function has the form
\begin{align}
&\frac{\rho^{\rm CR}(\Omega)}{\rho}
%=\\
%\nonumber
%&
=\frac {\re^{
\frac{3}{2}SM^*\cos^2\theta}\left(1+\beta^*-\frac{1}{5}\beta^*\sum\limits_m
\vert Y_{2m}(\Omega)\vert^2\,\sqrt{1-\frac{1}{5}\langle Y_{2m}^2(\Omega)\rangle_{\Omega}
M^*}\right)}{\int
\rd\Omega\, \re^{
\frac{3}{2}M^*S\cos^2\theta}\left(1+\beta^*-\frac{1}{5}\beta^*\sum\limits_m
\vert Y_{2m}(\Omega)\vert^2\,\sqrt{1-\frac{1}{5}\langle Y_{2m}^2(\Omega)\rangle_{\Omega}
M^*}\right)}\,.
\end{align}
We can also approximate the corrected single-particle distribution function in an exponential
form as
\begin{align}
\label{exp}
&\frac{\rho^{\rm EXP}(\Omega)}{\rho}=\frac{1}{Z''}
\exp\left[\frac{3}{2}M^*S\cos^2\theta-\frac{1}{5}\beta^*\sum\limits_{m} \vert
Y_{2m}(\Omega)\vert^2\,\sqrt{1-\frac{1}{5}\langle
Y_{2m}^2(\Omega)\rangle_{\Omega} M^*}\right],
\end{align}
where $Z''$ is the normalization constant such that
$\int\rho^{\rm EXP}(\Omega)\rd\Omega=\rho$.

Note that
\begin{align}
\vert
Y_{2m}(\Omega)\vert^2=\sum\limits_l \frac{5}{(2l+1)^{\frac{
1}{2}}}
\left(\begin{array}{ccc} {\,2}\quad\,{\,\,2}\quad{\,l}\\
{m}\,\,{-m}\,\,{\,\,0}\end{array}\right)\left(\begin{array}{ccc}{2}\quad{2}\quad{l}\\
{0}\quad{0}\quad{0}\end{array}\right)Y_{l0}(\Omega),
\end{align}
where $l=0,2,4$; $\left(\begin{smallmatrix} 2&2&l\\
m&-m&0\end{smallmatrix}\right)$ and $\left(\begin{smallmatrix} 2&2&l\\
0&0&0\end{smallmatrix}\right)$ are the corresponding
Clebsch-Gordon coefficients~\cite{gra}.

We can see that in the Gaussian approximation the dependence of
the single-particle distribution function on $\beta^*$ and $\rho^*$ is more complicated than in
the MFA: $\rho(\Omega)$ now depends not only on $M^*$ but there
is also a direct $\beta^*$-dependence and a $\langle
Y_{2m}^2(\Omega)\rangle_{\Omega}$-dependence. We also see that
in the Gaussian approximation the single-particle distribution function contains Legendre
polynomials of the second and fourth orders of molecule
orientations whereas in the linear approximation only Legendre
polynomials of the second order are present. From expression~(\ref{exp})
it is readily seen that the role of the fluctuation
term increases with an increase of inverse temperature
$\beta^*$.
%\begin{figure}
%\centerline{\includegraphics[scale=0.45]{density_correction}}
%\caption{Dependence of density on particle orientation $\cos\theta$ at
%$|\mathrm{M}^*| = 4.587$, $\mathrm{S} = 0.435$, $\beta^*= 0.19$.}
%\label{density_correction}
%\centerline{\includegraphics[scale=0.45]{density_correction_added}}
%\caption{Dependence of density on particle orientation $\cos\theta$ at
%$|\mathrm{M}^*| = 4.67$, $\mathrm{S} = 0.515$, $\beta^*= 1.40$.}
%\label{density_correction_added}
%\end{figure}
%\begin{figure}[htp]
 %\begin{center}
 %\subfigure[$|\mathrm{M}^*| = 4.587$,  $\mathrm{S} = 0.435$,  $\beta^*= 0.19$.]{\label{fig:edge-a}
    %\includegraphics[scale=0.45]{density_correction}}
   %\subfigure[$|\mathrm{M}^*| = 4.67$,  $\mathrm{S} = 0.515$,  $\beta^*= 1.40$.]{\label{fig:edge-b}
    %\includegraphics[scale=0.45]{density_correction_added}}
  %\end{center}
 %\caption{Dependence of density on particle orientation $\cos\theta$ at different values of parameters
  %$|\mathrm{M}^*|$,  $\mathrm{S}$,  $\beta^*$.}
 %\label{density_correction}
%\end{figure}
\subsection{Free energy, pressure, and chemical potential} For a
homogeneous system, the part of the Helmholtz free energy
responsible for field interaction can be calculated by
integrating with respect to the coupling parameter $\lambda$:
\begin{align}
\label{F}
F-F_{\rm id}=\frac{V}{2}\int\rd\mathbf{r}\rd\Omega_{1}\rd\Omega_{2}\rho(\Omega_{1})
\rho(\Omega_{2})\nu(r,\Omega_1\Omega_2)
\int\limits_{0}^{1}\rd\lambda[1+h(\lambda,r,\Omega_{1}\Omega_{2})],
\end{align}
where
\begin{align}
h(\lambda,r,\Omega_{1}\Omega_{2})&=\sum\limits_{m}h_{2m}(\lambda,r)
Y^*_{2m}(\Omega_1)Y_{2m}(\Omega_2)\nonumber\\
&=\sum\limits_{m}\left(-\frac{\lambda \beta {A}}{r}\frac{1}{5}
\exp\left[-r\sqrt{{\alpha^2+\langle{Y_{2m}^{2}(\Omega)}
\rangle_{\Omega}{\frac{1}{5}4\pi\rho\beta {A}\lambda}}}\right]\right)Y_{2m}^{*}(\Omega_1)Y_{2m}(\Omega_2).
\end{align}
Expression (\ref{F}) in terms of parameter
$B=\frac{1}{5}4\pi\rho\beta  A$ yields
\begin{align}
\label{freeenergy}
\beta\left(F-F_{\rm id}\right)&=V\frac{B\rho}{2\alpha^2}\langle Y_{20}(\Omega)\rangle^2_{\Omega}\nonumber\\
&+ V\sum\limits_{m}\left[
-\frac{\left(\alpha^{2}+\langle
Y_{2m}^2(\Omega)\rangle_{\Omega}B\right)^{3/2}}{12\pi}+\frac{\alpha^3}{12\pi}+
\frac{\alpha\langle{Y_{2m}^2(\Omega)}\rangle_{\Omega}
B}{8\pi}\right].
\end{align}
Having an explicit expression for the free energy, we can find
the pressure:
\begin{align}
\label{pressure}
 \beta P=&-\beta\left[\frac{\partial}{\partial V}F\right]_{T,N}=\rho+
 \frac{B\rho}{2\alpha^2}\left(\langle Y_{20}(\Omega)\rangle^2_{\Omega}+\rho\frac{\partial}{\partial\rho}
 \langle Y_{20}(\Omega)\rangle^2_{\Omega}\right)\nonumber\\
 &-\sum\limits_{m}\left(
-\frac{\left(\alpha^{2}+\langle
Y_{2m}^2(\Omega)\rangle_{\Omega}B\right)^{3/2}}{12\pi}+\frac{\alpha^3}{12\pi}+
\frac{\alpha\langle{Y_{2m}^2(\Omega)}\rangle_{\Omega}
B}{8\pi}\right)\nonumber\\
&+\frac{\rho B}{8\pi}\sum\limits_{m}\left (\frac{1}{\rho}\langle Y_{2m}^2(\Omega)\rangle_{\Omega}+
\frac{\partial}{\partial\rho}\langle
Y_{2m}^2(\Omega)\rangle_{\Omega}\right)\left(\alpha-\left(\alpha^2+\langle
Y_{2m}^2(\Omega)\rangle_{\Omega}B\right)^{1/2}\right),
\end{align}
where the derivatives of the averages $\langle
\ldots\rangle_{\Omega}$ are equal to
\begin{align}
\frac{\partial}{\partial\rho}\left\langle
Y_{2m}^2(\Omega)\right\rangle_{\Omega}&=\frac{B}{\rho\alpha^2}\left(\left\langle
Y_{20}(\Omega)\right\rangle_{\Omega}+\rho\frac{\partial}{\partial\rho}\left\langle
Y_{20}(\Omega)\right\rangle_{\Omega}\right)\nonumber\\
&\times\bigg(\left\langle
Y_{2m}^2(\Omega)\right\rangle_{\Omega}\left\langle
Y_{20}(\Omega)\right\rangle_{\Omega}-\left\langle Y_{2m}^2(\Omega)Y_{20}(\Omega)\right\rangle_{\Omega}\bigg),\\
\frac{\partial}{\partial\rho}\left\langle
Y_{20}(\Omega)\right\rangle_{\Omega}&=\frac{  B\left[\left\langle
Y_{20}(\Omega)\right\rangle_{\Omega}^3-\left\langle
Y_{20}^2(\Omega)\right\rangle_{\Omega}\left\langle
Y_{20}(\Omega)\right\rangle_{\Omega}\right]}{ \rho\alpha^2-\rho B\left[\left\langle
Y_{20}(\Omega)\right\rangle^2_{\Omega}-\left\langle
Y_{20}^2(\Omega)\right\rangle_{\Omega}\right]}\,.
\end{align}
In the isotropic phase $\langle
Y_{20}(\Omega)\rangle_{\Omega}=0$ and expression
(\ref{pressure}) considerably simplifies:
\begin{align}
\label{{pressure},{chemical}}
\beta P&=- \beta\left [\frac{\partial}{\partial V}F\right]_{T,N}=
   \rho+\sum\limits_{m}\left(
\frac{\alpha^{2}\left(\alpha^{2}+B_{2m}\right)^{1/2} }{12\pi}
-\frac{\alpha^3}{12\pi}
- \frac{B_{2m}\left(\alpha^{2}+B_{2m}\right)^{1/2}}{24\pi} \right),
%\beta P&=-\left[\frac{\partial}{\partial V}F\right]_{T,N}=\rho-\sum\limits_{m}\left(
%-\frac{\left(\alpha^{2}+\langle
%Y_{2m}^2(\Omega)\rangle_{\Omega}B\right)^{3/2}}{12\pi}+\frac{\alpha^3}{12\pi}+
%\frac{\alpha\langle{Y_{2m}^2(\Omega)}\rangle_{\Omega}
%B}{8\pi}\right)\\
%&+\frac{B}{8\pi}\sum\limits_{m} \langle Y_{2m}^2(\Omega)\rangle_{\Omega}
%\left(\alpha-\left(\alpha^2+\langle
%Y_{2m}^2(\Omega)\rangle_{\Omega}B\right)^{1/2}\right).
%\nonumber
\end{align}
where $B_{2m}\equiv \langle Y_{2m}^2(\Omega)\rangle_{\Omega}B$.
This expression is similar to the one obtained in~\cite{soviak,molphys}, where there is a supplementary summation over
$m$ and the quantity $B$ is replaced by $B_{2m}$\,.
The chemical potential $\mu$ of the fluid can be found from
expressions~(\ref{freeenergy}) and~(\ref{pressure}) as
$\mu=\left(F+PV\right)/N$ and equals
\begin{align}
\label{chemical}
\beta\mu&=\ln\left(\rho\Lambda_{\rm T}^3\Lambda_{\rm R}\right)+\frac{B}{\alpha^2}
\left(\langle{Y_{20}(\Omega)}\rangle^2_{\Omega}+\frac{1}{2}\rho\frac{\partial}{\partial\rho}\,
\langle{Y_{20}(\Omega)}\rangle^2_{\Omega}\right)\nonumber\\
&+\frac{B}{8\pi}\sum\limits_{m}\left (\frac{1}{\rho}\langle Y_{2m}^2(\Omega)\rangle_{\Omega}+
\frac{\partial}{\partial\rho}\langle
Y_{2m}^2(\Omega)\rangle_{\Omega}\right)\left(\alpha-\left(\alpha^2+\langle
Y_{2m}^2(\Omega)\rangle_{\Omega}B\right)^{1/2}\right).
\end{align}
\subsection{Broken symmetry problem and the elasticity
constant} A specific feature of the considered molecular fluid
is a broken symmetry which appears in the absence of an orienting
external field. In~\cite{hoso,hoso1,soho} using the
Lovett-Mou-Buff-Wertheim equation~\cite{lomo,we} an exact
relation for orientationally non-uniform fluids was obtained:
\begin{align}
\label{tlmbw}
\pmb{\nabla}_{\Omega_1}\ln\rho(\Omega_1)=\int
\rd\mathbf{r}_{12}\rd\Omega_2C(\mathbf{r}_{12}\,,\Omega_1\Omega_2)
\pmb{\nabla}_{\Omega_2}\rho(\Omega_2),
\end{align}
where $C(\mathbf{r}_{12}\,,\Omega_1\Omega_2)$ is the direct
correlation function which in the RPA is given by equation~(\ref{closure}).

Equation (\ref{tlmbw}) is also known as the
integro-differential form of the Ward identity~\cite{hoso,soho,wt}. The angular gradient operator
$\pmb{\nabla}_{\Omega}$ decomposes into 3 spherical components
${\nabla}_{0}$\,, ${\nabla}_{+}$\,, ${\nabla}_{-}$\,. As
${\nabla}_{0}$ is oriented in the direction of liquid crystal,
it vanishes due to rotational invariance. For other components,
the following relations hold~\cite{gra}:
\begin{align}
\nabla_{\pm}Y_{lm}(\Omega)=\left[l(l+1)-m(m\pm
1)\right]^{1/2}Y_{l,m\pm 1}\,.
\end{align}
The direct correlation function can be expanded as~\cite{hoso}
\begin{align}
C(r_{12}\,,\Omega_1\Omega_2)=\sum\limits_{lml'm'}C_{lml'm'}(r_{12})Y_{lm}^*(\Omega_1)Y_{l'm'}(\Omega_2).
\end{align}
Due to the axial symmetry of a nematic, $m=m'$. From~(\ref{tlmbw}) we derive the following relation for the
average $\langle Y_{21}^2(\Omega)\rangle_\Omega$\,:
\begin{align}
\label{ward}
C_{221}\rho\langle Y_{21}^2(\Omega)\rangle_\Omega=1,
\end{align}
where
\begin{align}
\label{ward1}
C_{221}=\int \rd\mathbf{r}_{12}C_{221}(r_{12})=-\frac{1}{5}\frac{4\pi\beta A}{\alpha^2}\,.
\end{align}
Due to condition~(\ref{ward}), harmonic $h_{221}(k)$ diverges
in the limit $\mathbf{k}\rightarrow 0$ signalling the
appearance of Goldstone modes in the system. This phenomenon is
responsible for a number of unique properties of nematics such
as elastic behavior and critical light scattering. The other
harmonics of the pair correlation function should be finite
according to the phenomenological theory of de Gennes~\cite{ge}. In the considered approximation~(\ref{closure}) we
have
\begin{align}
\label{simple}
C_{220}=C_{221}=C_{222}
\end{align}
and due to (\ref{ward}) and (\ref{ward1})
\begin{align}
\left(\alpha^2+\frac{1}{5}4\pi\rho\beta A \left\langle Y_{2m}^2(\Omega)\right\rangle_{\Omega}\right)^{1/2}=
\alpha\left(1-\frac{\left\langle Y_{2m}^2(\Omega)\right\rangle_{\Omega}}
{\left\langle Y_{21}^2(\Omega)\right\rangle_{\Omega}}\right)^{1/2}.
\end{align}
As we can see from figure~\ref{buka3}, for $m=2$, the ratio
$\left\langle
Y_{22}^2(\Omega)\right\rangle_{\Omega}/\left\langle
Y_{21}^2(\Omega)\right\rangle_{\Omega}<1$ and $h_{222}(r)$ is
well defined. However, for $m=0$, the ratio $\left\langle
Y_{20}^2(\Omega)\right\rangle_{\Omega}/\left\langle
Y_{21}^2(\Omega)\right\rangle_{\Omega}>1$ and $h_{220}(r)$ is
not defined. For this reason, a problem arises also for
thermodynamic properties and for the density correction. This
problem is connected with approximation~(\ref{closure}) which
leads to~(\ref{simple}). We should mention that in previous
works~\cite{hoso,hoso1,soho} carried out in the mean spherical
approximation equation~(\ref{simple}) is not true and the problem being discussed
 does not appear. We think that for point particles we
can also solve this problem by introducing an additional
isotropic interaction. We will consider this aspect of the
problem in a separate paper.

Broken symmetry is connected with elasticity properties of the
considered fluid which are described by the elasticity constant
$K$.  Formal expression for $K$ as proposed by Poniewiersky and
Stecki~\cite{post} within our model reads
\begin{align}
\label{el1}
\beta K&=\frac{1}{6}\int \rd\mathbf{r}\rd\Omega_1 \rd\Omega_2 r^2 \frac{\partial \rho(\Omega_1)}{\partial \cos\theta_1}
\frac{\partial \rho(\Omega_2)}{\partial \cos\theta_2}n_x(\Omega_1)n_x(\Omega_2)C(r,\Omega_1\Omega_2)\nonumber\\
&=10\pi\rho^2 S^2\int r^4 C_{221}(r)\rd r=-12\pi\rho^2 S^2\frac{\beta A}{\alpha^4}\,.
\end{align}
Another way to calculate elastic constants comes from the
theory of hydrodynamic fluctuations~\cite{post} and can be
found as
\begin{equation}
\label{el2}
\beta K=3\left[\lim\limits_{k\rightarrow 0}k^2
h_{221}(k)\frac{\langle Y^2_{21}(\Omega)\rangle^2_{\Omega}}{\langle Y_{20}(\Omega)\rangle^2_{\Omega}}\right]^{-1}=
-12\pi\rho^2 S^2\frac{\beta A}{\alpha^4}\,.
\end{equation}
As we see, both expressions for $K$ are identical.

\begin{figure}[!t]
\vspace{-0.5cm}
\hspace{-1cm}
\includegraphics[width=0.6\textwidth]{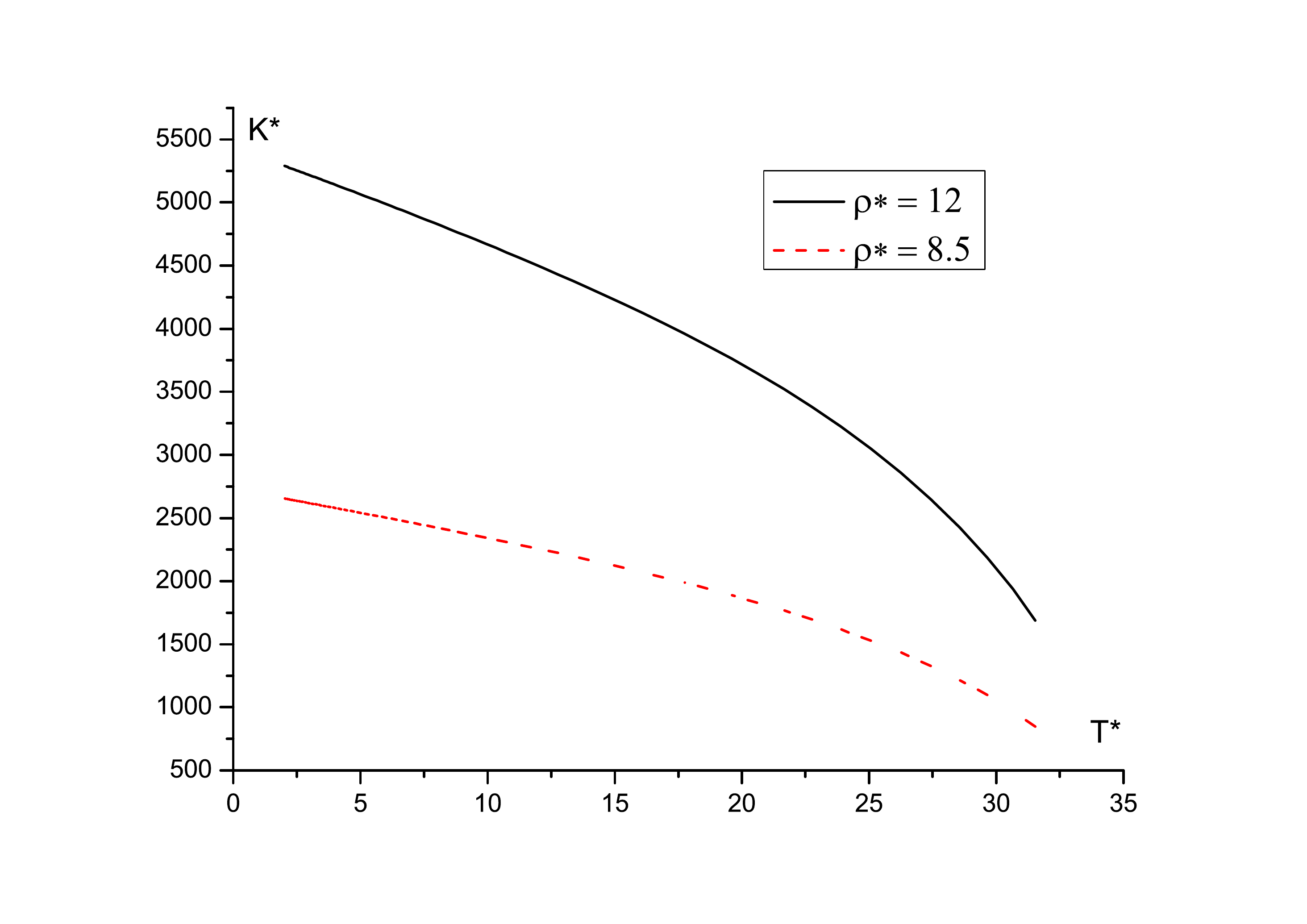}%
\hfill%
\hspace{-1.5cm}
\includegraphics[width=0.6\textwidth]{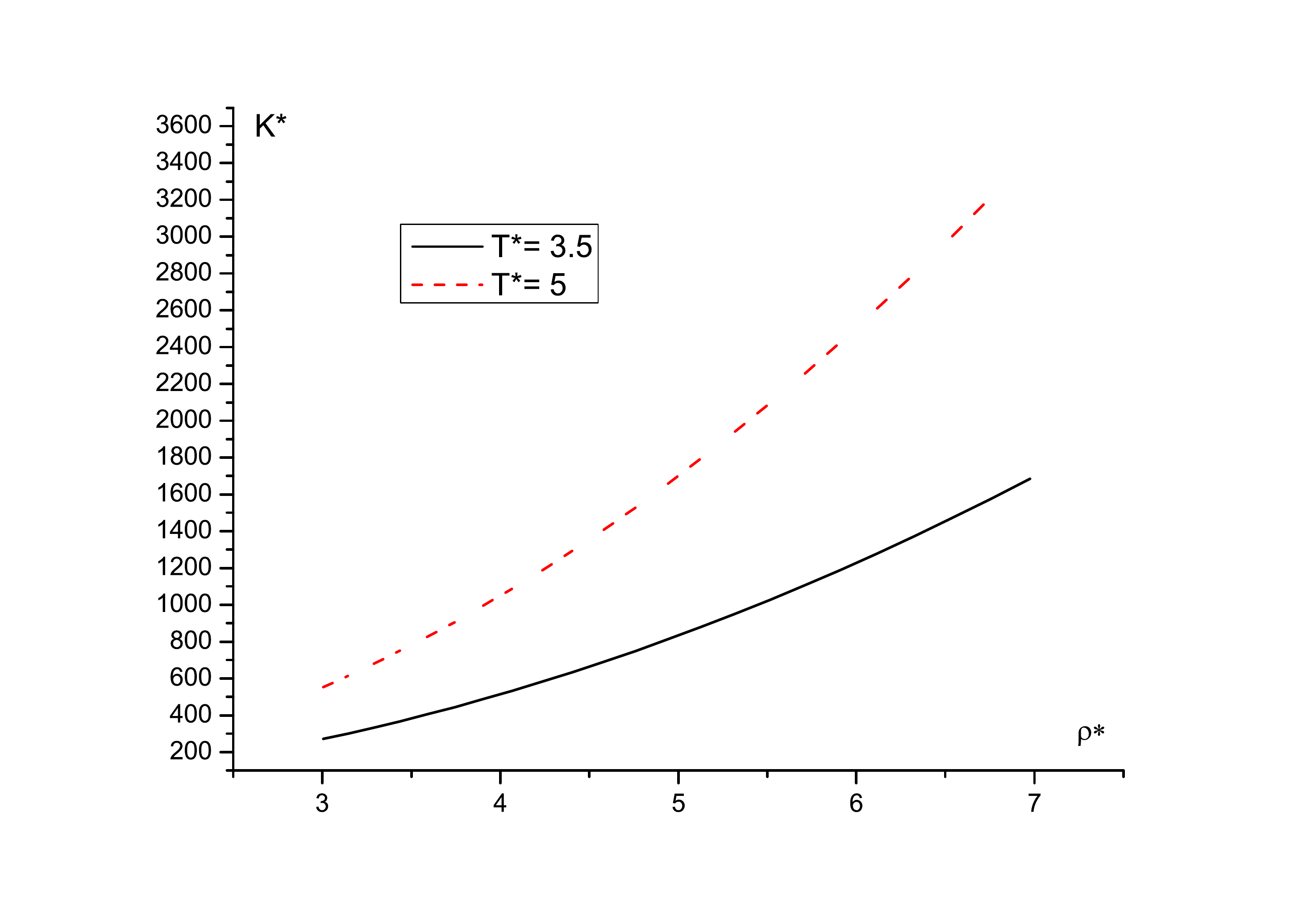}%
\vspace{-1cm}
\\
\parbox[t]{0.5\textwidth}{%
\caption{Temperature dependence of the elasticity constant.}
\label{4a}
}%
\hfill%
\parbox[t]{0.5\textwidth}{%
\caption{Density dependence of the elasticity constant.}
\label{4b}
}%
\end{figure}

%\begin{figure}
%\centerline{\includegraphics[scale=0.45]{elasticity_on_temperature}}
%\caption{Temperature dependence of the elasticity constant}
%\label{4a}
%\center\includegraphics[scale=0.45]{elasticity_on_density}
%\caption{Density dependence of the elasticity constant}
%\label{4b}
%\end{figure}
Dependencies of the reduced elasticity constant
$K^*=-K/A\alpha^2=12\pi S^2\rho^{*2}$ on temperature $T^*$ and
density $\rho^*$ are presented in figures~\ref{4a} and~\ref{4b}. One can see that with increasing temperature the
elasticity constant decreases and with increasing density it
increases. The effect is more pronounced respectively for
larger density and for larger temperature. This result is in
agreement with the previous result~\cite{soho} obtained in the
framework of the mean spherical approximation for a non-point
model of the nematogenic Maier-Saupe fluid.

\section{Conclusions}
In this paper, for the first time the field theoretical
approach is applied to the description of correlation functions
and thermodynamic properties of molecular anisotropic fluids.
As an example we consider a Maier-Saupe nematogenic fluid with
the Yukawa potential of interparticle interaction. By expanding
the Hamiltonian in powers of density fluctuations we examine
the system in the mean field and Gaussian approximations.

In the mean field approximation, we obtain analytical
expressions for the single-particle distribution function and the orientational order
parameter which are in agreement with the classical Maier-Saupe
theory for nematic ordered fluids.

In the Gaussian approximation we find analytical expressions
for the pair correlation function, the free energy, the
pressure, the chemical potential, and the elasticity constant.
We calculate the correction to the mean field single-particle distribution function due to
fluctuations and show that the corrected single-particle distribution function has a more
complex dependence on density and temperature compared to the
MFA. In contrast to the MFA single-particle distribution function, it includes the fourth
order Legendre polynomials of molecule orientations in addition
to the second order ones. We also use Ward symmetry identity to
derive a simple expression for the average of spherical
harmonic $\left\langle Y_{21}^2(\Omega)\right\rangle$. One
consequence of this condition is that for the system to be
stable, the distance-dependent part of the interaction should be
attractive. We show that in the Gaussian approximation the harmonic
$h_{221}(k)$ diverges in the limit $\mathbf{k}\rightarrow 0$.
Such a situation occurs at the phase transition from an
isotropic to a nematic phase. This change in symmetry causes
collective fluctuations known as the Goldstone modes. However,
in the RPA for the considered system, the harmonic $h_{220}(k)$ is
not defined. We hope to solve this problem in our next paper by
introducing additional isotropic interparticle interactions.

\section*{Acknowledgements}
M. Holovko and D. di Caprio are grateful for the support to the
National Academy of Sciences of Ukraine (NASU) and to the Centre
National de la Recherche Scientifique (CNRS) (project no.~21303), I.~Kravtsiv is grateful for the support to the French
embassy in Ukraine (the grant of the French government for PhD
programs in partnership). The authors also thank Dr.~O.V.~Patsahan for the useful comments.

\newpage

\ukrainianpart

\title{Нематичний плин Майєра-Заупе: теоретико-польовий підхід}
\author{М. Головко\refaddr{label1}, Д. ді Капріо\refaddr{label2}, І. Кравців\refaddr{label1}}
\addresses{
\addr{label1} Інститут фізики конденсованих систем НАН України,  вул. І. Свєнціцького, 1, 79011 Львів, Україна
\addr{label2} Лабораторія електрохімії, хімії поверхонь і енергетичного моделювання, \\
Відділення хімії вищої національної школи ПаріТех, пл.~Жуссю, 4, 75005~Париж, Франція
}

\makeukrtitle

\begin{abstract}
\tolerance=3000%

Ми застосовуємо теоретико-польовий підхід для вивчення структурних і термодинамічних властивостей однорідного нематичного плину Майєра-Заупе з анізотропною взаємодією типу Юкави. У наближенні середнього поля нами отримано стандартну теорію Майєра-Заупе для рідких кристалів. У цій теорії одночастинкова функція розподілу виражається через поліном Лєжандра другого порядку взаємної орієнтації частинок. У гаусівському наближенні нами отримано аналітичні вирази для кореляційних функцій, константи еластичності, вільної енергії, тиску і хімічного потенціалу. За допомогою тотожності Ворда нами встановлено просту умову для кореляційних функцій. Нами також знайдено поправки внаслідок флуктуацій і показано що вираз для одночастинкової функції розподілу вже містить поліноми Лєжандра вищих порядків.
\keywords нематичний плин Майєра-Заупе, теоретико-польовий підхід, кореляційна функція, термодинаміка

\end{abstract}

\end{document}